%% file: 00_Main.tex
\documentclass[11pt]{article}
\usepackage{geometry}
\usepackage{graphicx}
\graphicspath{{figures/}}

\usepackage{authblk}

\usepackage{fancyhdr}
\usepackage[yyyymmdd]{datetime}

\usepackage{indentfirst}
\usepackage{relsize}
\usepackage{amsmath}
\usepackage{tabularx,booktabs}
\newcolumntype{C}{>{\centering\arraybackslash} X }
\usepackage{array}
\usepackage{multirow}
\usepackage{listings}
\usepackage{ulem}
\usepackage{verbatim}
\usepackage{setspace}
\usepackage{subfiles} 

\usepackage{xcolor}
\definecolor{codegreen}{rgb}{0,0.6,0}
\definecolor{codegray}{rgb}{0.5,0.5,0.5}
\definecolor{codepurple}{rgb}{0.58,0,0.82}
\definecolor{backcolour}{rgb}{0.95,0.95,0.92}

\lstdefinestyle{mystyle}{
    backgroundcolor=\color{backcolour},   
    commentstyle=\color{codegreen},
    keywordstyle=\color{magenta},
    numberstyle=\tiny\color{codegray},
    stringstyle=\color{codepurple},
    basicstyle={\fontsize{9pt}{10.8pt}\ttfamily},
    breakatwhitespace=false,         
    breaklines=true,                 
    captionpos=b,                    
    keepspaces=true,                 
    numbers=left,                    
    numbersep=5pt,                  
    showspaces=false,                
    showstringspaces=false,
    showtabs=false,                  
    tabsize=4
}
\usepackage[hidelinks]{hyperref}

\makeatletter
\AtBeginDocument{%
  \def\doi#1{\url{https://doi.org/#1}}}
\makeatother

\title{Influence zones for continuous beam systems}

\author[1,*]{Adrien Gallet}
\author[1]{Andrew Liew}
\author[1]{Iman Hajirasouliha}
\author[2]{Danny Smyl}

\affil[1]{Department of Civil and Structural Engineering, University of Sheffield, Sheffield UK}
\affil[2]{Department of Civil, Coastal, and Environmental Engineering, University of South Alabama, Mobile, AL, USA}
\affil[ ]{ } 
\affil[*]{Author for correspondence: \textit{agallet1@sheffield.ac.uk}}
\affil[ ]{ } 
\affil[ ]{Version: Author's Original (AO)}

\date{2023--02--23} 

\begin{document}
    
\maketitle 

\begin{abstract}
    \noindent Unlike influence lines, the concept of influence zones is remarkably absent within the field of structural engineering, despite its existence in the closely related domain of geotechnics. This paper proposes the novel concept of a structural influence zone in relation to continuous beam systems and explores its size numerically with various design constraints applicable to steel framed buildings. The key challenge involves explicitly defining the critical load arrangements, and is tackled by using the novel concepts of 
    polarity sequences and polarity zones. These lead to the identification of flexural and (discovery of) shear load arrangements, with an equation demarcating when the latter arises. After developing algorithms that help identify both types of critical load arrangements, design data sets are generated and the influence zone values are extracted. The results indicate that the influence zone under ultimate state considerations is typically less than 3, rising to a maximum size of 5 adjacent members for any given continuous beam. Additional insights from the influence zone concept, specifically in comparison to influence lines, are highlighted, and the avenues for future research, such as in relation to the newly identified shear load arrangements, are discussed.
\end{abstract}


\subfile{01_Introduction}

\subfile{02_Method}
\subfile{03_Critical_Load_Arrangements}
\subfile{04_Influence_Zones}
\subfile{05_Discussion}
\subfile{06_Conclusion}

\section*{Acknowledgements}
\vspace{0.5cm}

\noindent \textbf{Authors' contributions:} Adrien Gallet: Conceptualization, Methodology, Investigation, Software, Formal analysis, Validation, Visualization, Writing - Original Draft \hspace{0.5cm} Andrew Liew: Writing - Review \& Editing \hspace{0.5cm} Iman Hajirasouliha: Writing - Review \& Editing \hspace{0.5cm} Danny Smyl: Supervision, Writing - Review \& Editing

\vspace{0.5cm}

\noindent \textbf{Data statement:} Data used in this article are available from the authors upon request.
\vspace{0.5cm}

\noindent \textbf{Competing interests:} The authors declare that they have no competing interests.

\vspace{0.5cm}

\bibliographystyle{elsarticle-num}
\bibliography{07_References}

\clearpage
\appendix

\begin{minipage}{0.95\textwidth}
\section{Algorithm 1 - Flexural load arrangements}

\label{app:algorithm_1}
\begin{center}
\begin{lstlisting}[language=Python, caption=Flexural load arrangement algorithm in Python with both alternating and adjacent arrangements for a continuous beam system with $m$ members that creates set $\mathbf{J_{flex}}$ with O(m) time complexity., label=algo:flexural_set]
# Create the alternating load arrangements - altLoadArr
if m % 2 == 0: altLoadArr = [[1,0]*(m//2)]
else: altLoadArr = [[1,0]*(m//2) + [1]]

# Create the adjacent load arrangements - adjLoadArr
adjLoadArr = []
if m > 1:
    for i in range(m-1):
        # Create the start loadArr
        if i % 2 == 0: startLoadArr = [1,0]*(i//2)
        else: startLoadArr = [0,1]*(i//2) + [0]

        # Create the end loadArr
        if (m-i) % 2 == 0: endLoadArr = [0,1]*((m-i-2)//2)
        else: endLoadArr = [0,1]*((m-i-2)//2) + [0]

        # Append loadArr together with adjacent loaded spans
        adjLoadArr.append(startLoadArr + [1,1] + endLoadArr)

# Create positive J_flex load arrangements
J_flex_pos = altLoadArr + adjLoadArr

# Evaluate polar opposites - negative J_flex
J_flex_neg = []
for loadArr in J_flex_pos:
    J_flex_neg.append([1 if act == 0 else 0
                       for act in loadArr])

# Evaluate J_flex
J_flex = J_flex_pos + J_flex_neg
\end{lstlisting}
\end{center}
\end{minipage}

\begin{minipage}{0.95\textwidth}
\section{Algorithm 2 - Shear load arrangements}
\label{app:algorithm_2}
\begin{center}
\begin{lstlisting}[language=Python, caption=Shear load arrangement algorithm in Python to generate arrangements belonging to set $\mathbf{J_{shear}}$ based on a given flexural load arrangements \textit{loadArr}{,} the indices of susceptible \textit{shearBeams}{,} starting at beam index \textit{start}{,} for a system size with \textit{m} members. One single pass has a time complexity of O(m){,} yet generating the entire set $\mathbf{J_{shear}}$ is $O(m^2\, 2^n)$., label=algo:shear_set]
# Function to identify shear load arrangements
def shearLoadArr(loadArr: list, shearBeams: list, start: int):
    # Iterate in both directions
    for direction in [-1, 1]:
        # Establish while loop variables
        finishing = False; finished = False
        updating = False; i = start

        # Iterate through the beam system
        while finished == False:
            i = i + direction # Move to the next beam
            
            # Case 1: End of beam system is reached
            if i < 0 or i >= len(loadArr):
                finished = True
            
            # Case 2: No shear beam has been encountered yet
            elif updating == False and finishing == False:
                # Check if current beam is a shear beam
                if i in shearBeams:
                    updating = True; updateAct = loadArr[i]
            
            # Case 3: A shear beam has been encountered
            elif updating == True and finishing == False:
                # Update activation factor of current beam
                loadArr[i] = updateAct
                # Check if current beam is a shear beam
                if i not in shearBeams:
                    updating = False; finishing = True
            
            # Case 4: Alternate remaining activation factors
            elif finishing == True:
                loadArr[i] = (loadArr[i-direction] + 1) % 2
                
                # If another shear beam is encountered
                if i in shearBeams: 
                    updateAct = loadArr[i]
                    updating = True; finishing = False

    return loadArr
\end{lstlisting}
\end{center}
\end{minipage}

\end{document}

%% file: 01_Introduction.tex
\section{Introduction}
\label{sec:introduction}
Influence lines, which derive from Betti's theorem established in 1872 \cite{betti_teoria_1872}, are a well-established tool in structural engineering to identify the worst-case load placement on structural systems \cite{blake_civil_1989, karnovsky_advanced_2010, hibbeler_structural_2015}, and are widely applied in research related to continuous beam systems \cite{hosur_influence_1996, fiorillo_application_2015}, rigid frames \cite{buckley_basic_1997}, bridge engineering \cite{zheng_development_2019} and structural health monitoring \cite{zhu_structural_2015, chen_damage_2018}. Influence zones, on the other hand, also known as zones of influence, are an established concept within the field of geotechnical engineering, helping to identify the area of engineering soils likely to be affected by loading due to sub- and superstructure construction \cite{bobrowsky_zone_2018}, providing geotechnical engineers valuable design insight in deep foundation design \cite{yang_influence_2006, ekanayake_influence_2013}, settlement estimations \cite{kuklik_efficient_2004} and preserving groundwater supplies \cite{cao_determination_2021}.

Despite the obvious discipline link between geotechnical and structural engineering, the equivalent use of an influence zone in structural engineering does not exist in literature. Here, the term  \textit{structural influence zone} would refer to the zone in which applied forces, stiffness provisions and support conditions, or changes thereof, impact the design of the surrounding structural system. 

The dearth of literature on such an \textit{influence zone} is surprising. 
For instance, the concept of influence zones also exists outside of geotechnical literature.
Some examples are available in research related to the study of saltwater-freshwater interfaces \cite{jakovovic_saltwater_2016}, harmful emission concentrations at traffic intersections \cite{goel_zone_2015}, reverse \textit{k}-nearest neighbour algorithms \cite{alvin_influence_2021, cheema_efficiently_2012}, propagation path of surfaces waves \cite{yoshizawa_determination_2002} and ecological studies on below-ground plant competition \cite{casper_defining_2003}.

Furthermore, one can readily identify situations where knowledge of the \textit{influence zone} could be beneficial in design. For example, the size of the influence zone could allow an engineer to avoid the need to model an entire structure for the design of a single element whilst being confident that structural information outside the influence zone is irrelevant, with direct applications in multi-disciplinary projects \cite{schlueter_building_2009}. 
The impact of late design changes (due to changes in loading or structural provisions), which are known to cause significant time lags until the associated engineering analysis is completed \cite{sinclair_riba_2020}, could be more effectively addressed by knowing immediately the selection of members impacted by the said design change. Similarly, engineers are typically required to verify assumptions made in preliminary design \cite{mason_structural_2011}.
In such cases, the use of an influence zone-based approach could guide what information to incorporate when building an independent model of the design problem. In all of these scenarios, there is valuable design insight to be gained from the \textit{influence zone}.

This article aims to address the above mentioned knowledge gap by numerically introducing the concept of influence zones in relation to continuous beam systems. First, the theory and methodology for evaluating the influence zone will be introduced in section \ref{sec:method}, followed by a systematic analysis of critical load arrangements in section \ref{sec:critical_load_arrangements}. The explicit formulations of critical load arrangements allow for the efficient generation of design data sets and the evaluation of their respective influence zones in section \ref{sec:influence_zones}, the results of which are discussed in section \ref{sec:discussion}. In addition to the \textit{influence zone}, this paper proposes other novel concepts such as \textit{polarity zones}, identifies an entirely new set of critical pattern loads named \textit{shear load arrangements}, and proposes efficient \textit{load arrangement algorithms} for continuous beam systems of arbitrary member size.

%% file: 02_Method.tex
\section{Methodology}
\label{sec:method}
\subsection{Overview}

\label{sec:method_overview}
\begin{figure*}[!htb]
	\centering
    \includegraphics[width=\textwidth,height=\textheight,keepaspectratio]{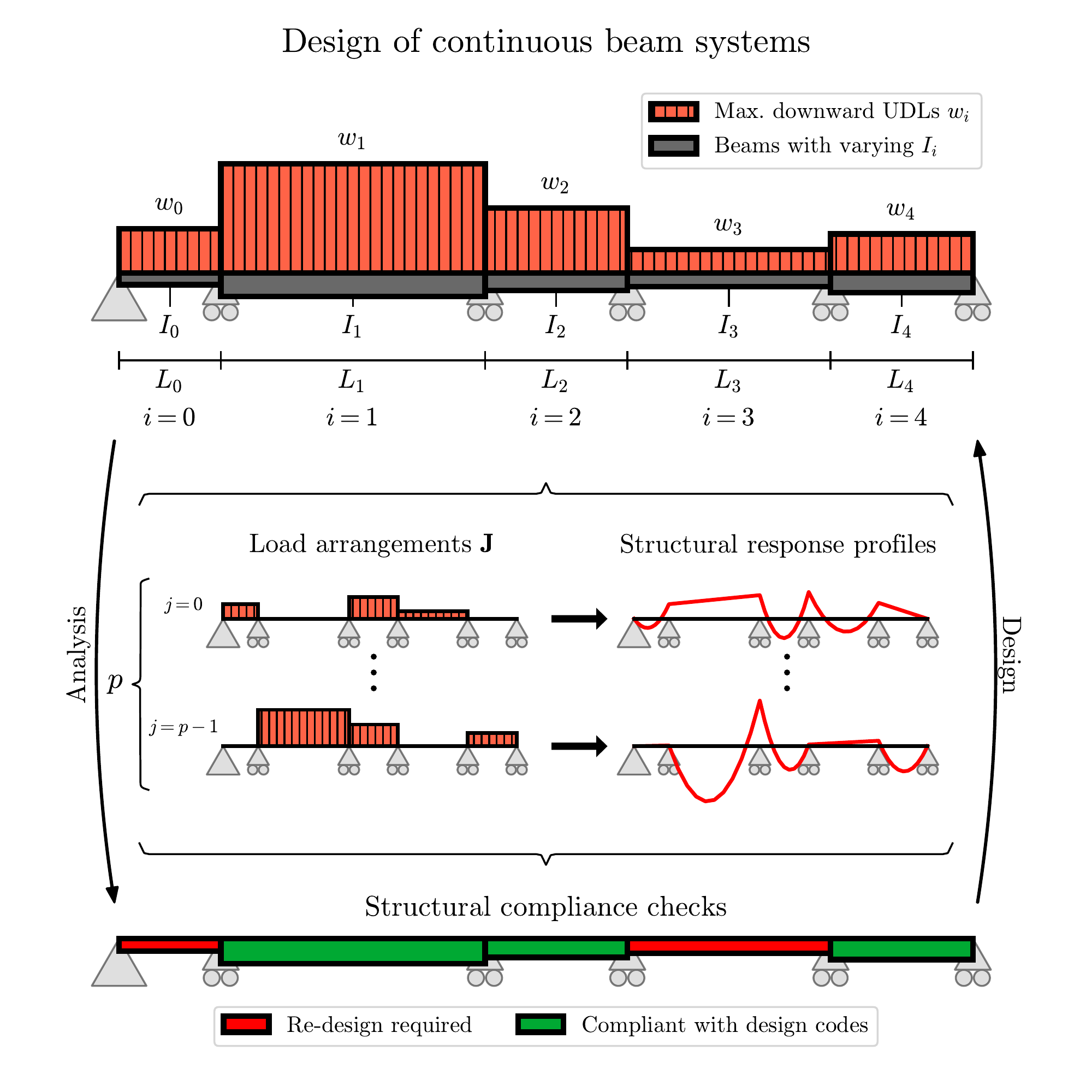}
	\caption{An exemplary continuous beam system with $m=5$ members, subjected to UDLs $\mathbf{w}$, spans $\mathbf{L}$ and with designed cross-sectional properties $\mathbf{I}$, all indexed by $i$. The system's indeterminacy requires an iterative design process against various load arrangements $\mathbf{J}$ of size $p$ indexed by $j$.}
	\label{fig:design_system_overview}
\end{figure*}

Consider a continuous beam system, as shown in Figure \ref{fig:design_system_overview}, consisting out of $m$ members, indexed by $i$, which is subjected to $w_i$ uniformly distributed loads (UDLs) from vector $\mathbf{w}$, with each member having span length $L_i$ from vector $\mathbf{L}$. When designing this system to identify the minimum required structural properties of the members (size optimisation) denoted $I_i$ to form vector $\mathbf{I}$, it will need to be designed against the worst-case load arrangement (also known as pattern load) from the set of load arrangements $\mathbf{J}$ of size $p$. The over-restrained nature of this structural system (a function of the support fixity and structural connectivity) renders the continuous beam indeterminate. This means that the performance of the system is a function of the structural properties which need to be evaluated, and generally makes the design process iterative. Literature has well established formulations to design such indeterminate systems \cite{saka_mathematical_2013}.

\subsection{Influence zone formulations}
\label{sec:influence_zone_formulation}
\begin{figure*}[htb]
	\centering
    \includegraphics[width=\textwidth,height=\textheight,keepaspectratio]{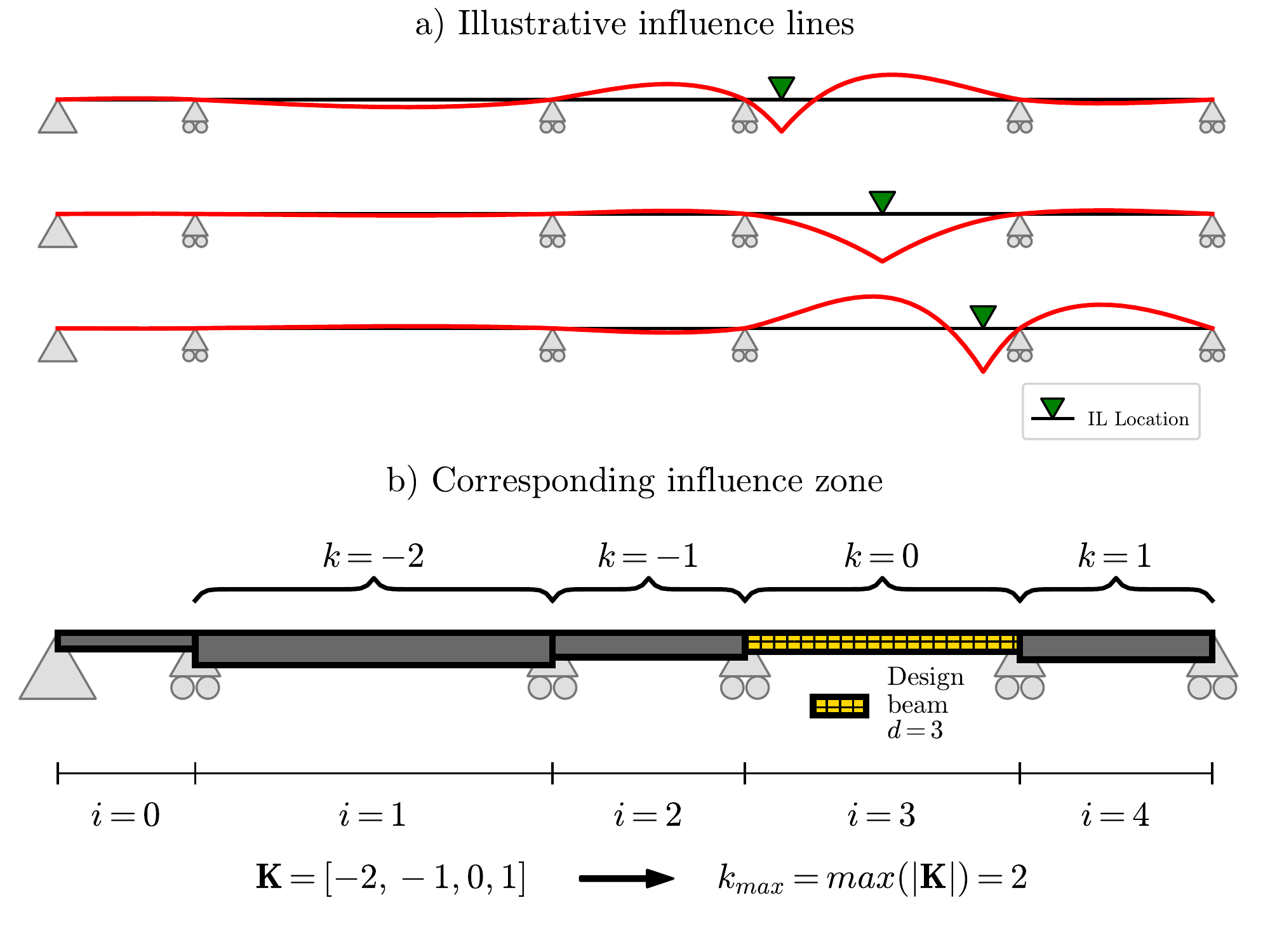}
	\caption{An example demonstrating how influence lines relate to influence zones, and what an influence zone of size $k_\mathrm{max}=2$ corresponds to in relation to a given design beam (here $d=3$, highlighted in yellow).}
	\label{fig:influence_zone_overview}
\end{figure*}

Suppose a member within a continuous beam system is designated as the \textit{design beam} by index $d$, and a discrete integer $k \in \mathbf{Z}$ indicates the index position of a member relative to the design beam at $d$. As shown in Figure \ref{fig:influence_zone_overview}, if $\mathbf{K}$ refers to the list of members identified in terms of $k$ that fall within the influence zone, then the size of the influence zone is denoted by $k_\mathrm{max}= \max (|\mathbf{K}|)$ with $k_\mathrm{max} \in \mathbf{N}^0$, representing the set of all positive integers and including 0. Two different formulations have been identified to evaluate the influence zone:

\begin{itemize}
    \item The \textit{local formulation} identifies the value of $k_\mathrm{max}$ based on whether the design information at the design beam $d$ significantly influences the surrounding members of indices $d-k_\mathrm{max} \le i \le d+k_\mathrm{max}$.
    \item The \textit{global formulation} identifies the value of $k_\mathrm{max}$ based on when the design information at members with indices $i<d-k_\mathrm{max}$ and $i>d+k_\mathrm{max}$ becomes inconsequential for the design beam at $d$.
\end{itemize}

For the continuous beam system established in Figure \ref{fig:design_system_overview}, the ``design information'' include the UDLs $w_i$ and spans $L_i$. Although the terms ``significantly influences'' and ``becomes inconsequential'' are currently undefined, they refer to an error threshold that will be explained later. Whilst the \textit{local} and \textit{global} formulations differ in terms of where the design information impact is measured from (locally at the design beam for the \textit{local formulation} or outside the influence zone for the \textit{global formulation}), as long as the design constraints are identical, the size of the influence zone $k_\mathrm{max}$ each formulation identifies will be identical.

There are various methodologies one could employ to establish the influence zone using either formulation.
For example, \textit{analytical} approaches making use of concepts such as perturbation theories based on the relationship between force vectors and stiffness matrices may be viable.
On the other hand, influence zones could be approached experimentally with the use of physical models or numerically with the use of finite element methods.

Each methodology has its own disadvantages. It is not intuitive how one would evaluate the size of the influence zone using a perturbation based approach if large design perturbations are required with multiple load arrangements. Experimental procedures would be limited by the number of design scenarios that can be tested, whilst numerical approaches would make mechanical assumptions on the material and structural behaviour of the system.

A numerical approach was preferred since it allows a multitude of design scenarios to be investigated and for statistical conclusions to be determined. Both the local and global formulations were attempted with the use of a numerical model, yet only the latter formulation was fully developed. This was because with the global formulation, the influence zone $k_\mathrm{max}$ could be measured in relation to the utilisation ratio of the design beam $d$ directly, which made evaluating and reporting the influence zone easier. In the local formulation, the utilisation ratio of all surroundings members outside the influence zone would have to be monitored.

\subsection{Mathematical formulation}
\label{sec:mathematical_formulation}
Mathematically, the global formulation can be expressed as follows. For a given continuous beam system as depicted in Figure \ref{fig:design_system_overview}, and the design constraints expressed in Equation \ref{eq:design_constraints},

\begin{equation}
    \label{eq:design_constraints}
    \begin{array}{r@{\ }c@{\ }l}
        w_\mathrm{min} & < w_i & < w_\mathrm{max} \\
        L_\mathrm{min} & < L_i & < L_\mathrm{max} \\
        I_\mathrm{min} & < I_i & < I_\mathrm{max}
    \end{array}
\end{equation}

\noindent the size of the influence zone of a given design beam $d$ is found when the value of $k_\mathrm{max} \in \mathbf{N}^0 : k_\mathrm{max} \in [0,m]$ \textbf{and} all values larger than $k_\mathrm{max}$ fulfil the following condition:

\begin{equation}
    \begin{gathered}
        \label{eq:global_formulation}
        \left| \, 1 - \frac{u_{d,\mathrm{cap}}}{u_{d,\mathrm{true}}} \, \right| \le \epsilon_\mathrm{max} \\[0.4cm]
        u_{d,\mathrm{cap}} = \max \left( \ \mathlarger{\sum_{i=-k_\mathrm{max}}^{k_\mathrm{max}}} \mathbf{u}_{d,i,\,j}(\mathbf{w, L, I, J}) \ \right)
        \end{gathered}
\end{equation}

\noindent where $\epsilon_{\mathrm{max}}$ represents the maximum error threshold for the difference between $u_{d,\mathrm{cap}}$, the captured utilisation ratio of the design beam $d$ for a given value of $k_\mathrm{max}$, and $u_{d,\mathrm{true}}$, the true utilisation ratio of the design beam $d$ if the contribution of all members of the continuous beam system had been considered. $\mathbf{u}_{d,i,\,j}$ is the utilisation ratio contribution function towards the design beam $d$ by member $i$ based on the UDLs $\mathbf{w}$, spans $\mathbf{L}$, structural properties $\mathbf{I}$ and load arrangements $\mathbf{J}$ indexed by $j$.

The global formulation as written in Equation \ref{eq:global_formulation} measures the point at which the contributions outside of $k_\mathrm{max}$ ``becomes inconsequential'' by minimising the difference between $u_{d,cap}$ and $u_{d,true}$ based on $\epsilon_\mathrm{max}$. As $k_\mathrm{max}$ increases, the ratio $u_{d,cap}/u_{d,true}$ will approach unity, attaining unity if all structural members ($k_\mathrm{max} = m$) are considered within the influence zone. If the error threshold $\epsilon_\mathrm{max}$ is relaxed, an influence zone less than the total number of beam members $m$ can be found. The influence zone is therefore a heuristic measure based on an acceptable maximum error threshold $\epsilon_\mathrm{max}$.

The importance of the design constraints as specified by Equation \ref{eq:design_constraints} is that they allow for the statistical estimation of the maximum influence zone size based on the diversity of design information variation that can arise. The maximum influence zone value for a type of structural system should always be understood with explicit reference to the design constraints it was evaluated by.

\subsection{Design constraints and assumptions}
\label{sec:design_constraints}
The design constraints considered in this investigation were chosen for their relevance in the design of continuous steel framed buildings, which is reflected by the range of UDLs and spans of the design data sets. Four individual design scenarios are considered to study the influence zone in depth, with each set featuring an increasing variation in span lengths and applied loads, summarised in Table \ref{table:design_sets}. Length and UDL values are discretized in $0.5\text{\ m}$ and $5 \text{\ kN/m}$ increments respectively, and are drawn from a random uniform distribution.

\begin{table*}[htb]
	\centering
    \small
    \begin{tabular}{m{2.8cm} m{2.0cm} m{3.0cm} m{3.0cm}}
		\toprule
        \multirow{1}{=}{\centering Data set} &
        \multirow{1}{=}{\centering $G_{k,i} =$} &
        \multirow{1}{=}{\centering $Q_{k,i} \in$} &
        \multirow{1}{=}{\centering $L_i \in$} \\ \midrule
        
        \multirow{2}{=}{\centering Set 1 \\ \textit{Zero variation}} &
        \multirow{2}{=}{\centering $3.0 \text{\,kN/m} + \text{self-weight}$} &
        \multirow{2}{=}{\centering $a$ for all $i$, with $a \in[0 \text{\,kN/m}, 60 \text{\,kN/m} ]$} &
        \multirow{2}{=}{\centering $b$ for all $i$, with $b \in [1 \text{\,m}, 12 \text{\,m}]$} \\ \\ [0.4cm]
        
        \multirow{2}{=}{\centering Set 2 \\ \textit{Low variation}} &
        \multirow{2}{=}{\centering$3.0 \text{\,kN/m} + \text{self-weight}$} &
        \multirow{2}{=}{\centering $[20 \text{\,kN/m}, 40 \text{\,kN/m} ]$} &
        \multirow{2}{=}{\centering $[4 \text{\,m}, 8 \text{\,m} ]$} \\ \\ [0.4cm]

        \multirow{2}{=}{\centering Set 3 \\ \textit{Medium variation}} &
        \multirow{2}{=}{\centering $3.0 \text{\,kN/m} + \text{self-weight}$} &
        \multirow{2}{=}{\centering $[10 \text{\,kN/m}, 50 \text{\,kN/m} ]$} &
        \multirow{2}{=}{\centering $[2 \text{\,m}, 10 \text{\,m} ]$} \\ \\ [0.4cm]

        \multirow{2}{=}{\centering Set 4 \\ \textit{High variation}} &
        \multirow{2}{=}{\centering $3.0 \text{\,kN/m} + \text{self-weight}$} &
        \multirow{2}{=}{\centering $[0 \text{\,kN/m}, 60 \text{\,kN/m} ]$} &
        \multirow{2}{=}{\centering $[1 \text{\,m}, 12 \text{\,m} ]$} \\ \\ \bottomrule
        
    \end{tabular}
    
    \caption{Design constraints for various design scenarios used in this investigation based on Eurocode terminology, with $G_k$ and $Q_k$ being the characteristic permanent and variable actions. Increasing set numbers correspond to increasing design variation, a proxy for design complexity. Span and UDL values are discretized in $0.5\text{\ m}$ and $5 \text{\ kN/m}$ increments respectively, and will be drawn from a random uniform distribution.}
	\label{table:design_sets}
\end{table*}

Further design and modelling constraints/assumptions include restricting the cross-sectional properties to prismatic BS EN 10365:2017 UKB I-sections, designing for S355 steel with perfectly linear elastic behaviour using Timoshenko-Ehrenfest beam theory (yet the design was conducted using plastic section properties as allowed by EN 1993-1-1 5.4.2(2) \cite{CEN_2015_BS_EN_1993_1_1_2005_A1_2014}). It was assumed that all spans are laterally restrained (and hence not susceptible to lateral instability), with elements designed against EC3 ULS checks (and notably not SLS requirements) with EN 1990 Eq. 6.10 load combination factors \cite{CEN_2010_BS_EN_1990_2002_A1_2005}.

\subsection{The key challenge}
\label{sec:key_challenge}
The most important aspect for evaluating the influence zone using equation \ref{eq:global_formulation} is the member-based utilisation ratio contribution function $\mathbf{u_\mathrm{d,i,\,j}}$. Whilst the UDLs $\mathbf{w}$ and spans $\mathbf{L}$ are given, the critical load arrangement from set $\mathbf{J}$ which will determine the required structural properties $\mathbf{I}$ is unknown. Furthermore, the critical load arrangement for a design beam could also differ based on the assumed value of the influence zone size $k_\mathrm{max}$. 

One approach would be to use a naive, brute-force procedure to trial every possible load arrangement to create the set $\mathbf{J_\mathrm{naive}}$ with a corresponding set size of $p_{naive}=2^m$ for $\mathbf{J}$ in equation \ref{eq:global_formulation}. This is not an issue for systems with few members, but if larger systems with $m>10$ members need to be modelled to study the influence zone in depth, a brute-force approach becomes computationally expensive. The issue of computational cost in relation to critical load arrangements of large-scale systems is well acknowledged in literature, and various methodologies have been employed using probability \cite{shiryayev_interpolation_1992} and possibility theories \cite{zadeh_fuzzy_1965, zadeh_fuzzy_1999}. Among the latter, fuzzy sets using interval finite-element methods have been shown to be efficient and accurate \cite{koyluoglu_interval_1995, mullen_bounds_1999}.

However, whilst these interval-based methods are effective at evaluating the bounds (the worst case force/moment value) of the critical load arrangement, they do not in fact reveal what this load arrangement looks like. This is problematic for the evaluation of the influence zone, since Equation \ref{eq:global_formulation} relies on being able to identify this set $\mathbf{J}$ explicitly. Another approach would be to use the load arrangements prescribed by design manuals, yet these consist out of a heuristic set of load arrangements that are known to be non-conservative \cite{mullen_bounds_1999}.

Due to these limitations, a rigorous study is conducted to identify and validate the set of critical load arrangements \textit{a priori} for the design problem identified in Section \ref{sec:method_overview}, labelled $\mathbf{J_\mathrm{crit}}$. This will not only reduce the computational cost of both designing the members and evaluating their influence zone with Equation \ref{eq:global_formulation}; it will also highlight the relationship between influence lines and influence zones, providing an intuition on the size of the latter. This study is followed by the numerical generation of randomly distributed data sets based on the design constraints identified in Section \ref{sec:design_constraints}, allowing the evaluation and statistical analysis of the influence zone for various continuous beam systems.

%% file: 03_Critical_Load_Arrangements.tex
\section{Critical load arrangement investigation}
\label{sec:critical_load_arrangements}
\subsection{Polarity sequences}
\label{sec:polarity_diagrams}
\begin{figure*}[!htb]
	\centering
    \includegraphics[width=\textwidth,height=\textheight,keepaspectratio]{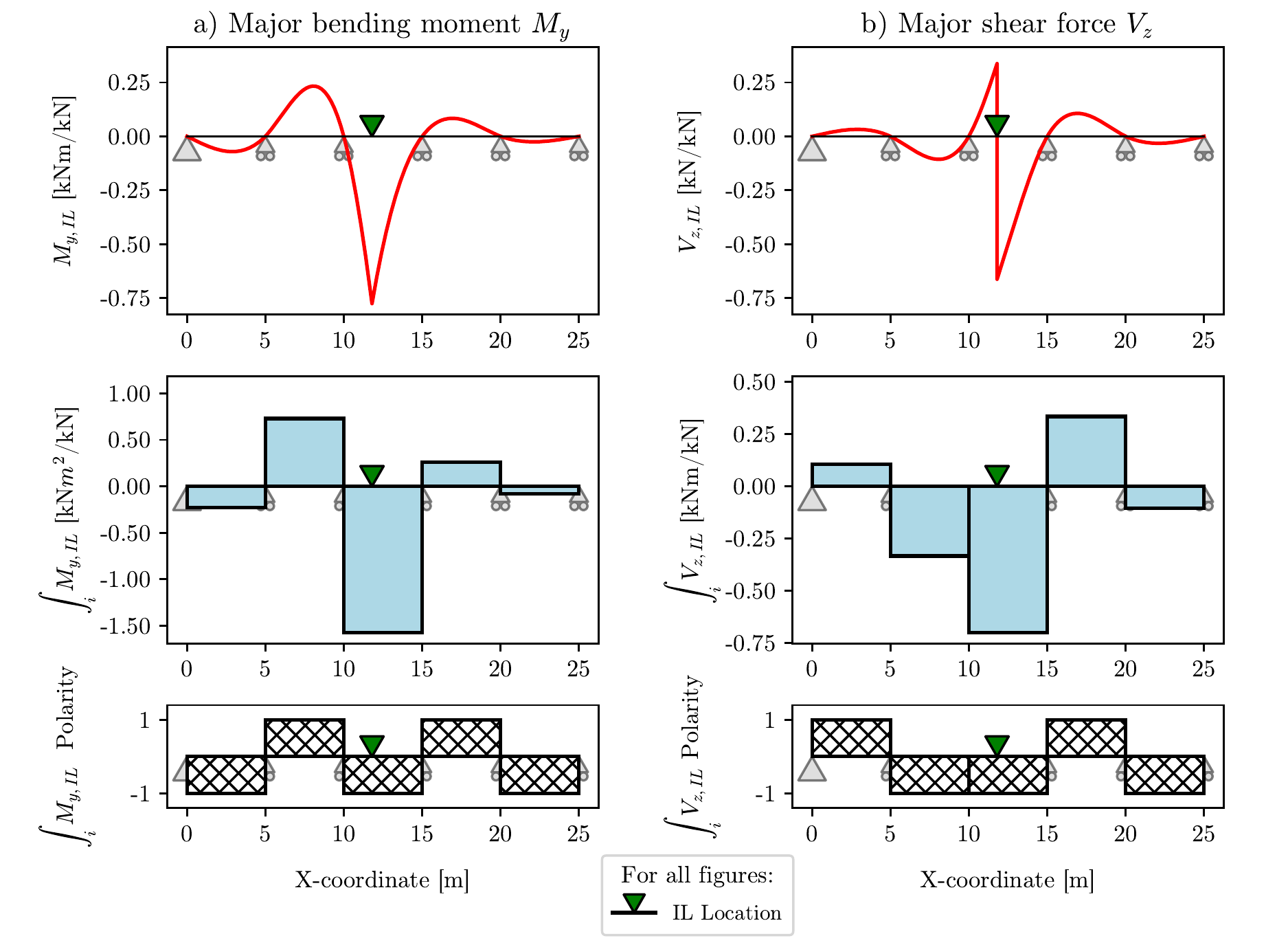}
	\caption{An exemplary process of arriving from influence line plots (top row) to polarity sequences (bottom row) via integrated influence lines (middle row) for a) major axis bending moment $M_y$ and b) major axis shear force $V_z$ about the specified influence line (IL) location.}
	\label{fig:influence_lines_to_polarity_zones}
\end{figure*}

Influence lines can be used to identify the critical load arrangements for a given continuous beam system. The design problem is restricted to positive UDL values only (no uplift) which can be activated on or off (1 or 0 as activation factors). Therefore, by integrating the influence line for each individual beam $i$, one can evaluate the net contribution (positive or negative) a given beam causes in terms of bending moments and shear forces at the influence line (IL) location when subjected to a positive, unit UDL. The net-contribution of each beam can be either positive or negative at the IL location, that is ``hogging or sagging'' for bending moments and ``clockwise or anti-clockwise'' for shear forces, respectively, which is termed as the \textit{polarity} of that particular beam.  This procedure is shown in Figure \ref{fig:influence_lines_to_polarity_zones}.

The last row of Figure \ref{fig:influence_lines_to_polarity_zones} therefore reflects a particular \textit{polarity sequence} for a given IL location, which can be directly used to identify the critical load arrangement for that IL location. When all beams of positive polarity are loaded, then the maximum positive internal forces are generated at the IL location, and vice-versa, loading the negative polarity members leads to the maximum negative internal forces.

\subsection{Polarity zones}
\label{sec:polarity_zones}
\begin{figure*}[!htb]
	\centering
    \includegraphics[width=0.99\textwidth,height=\textheight,keepaspectratio]{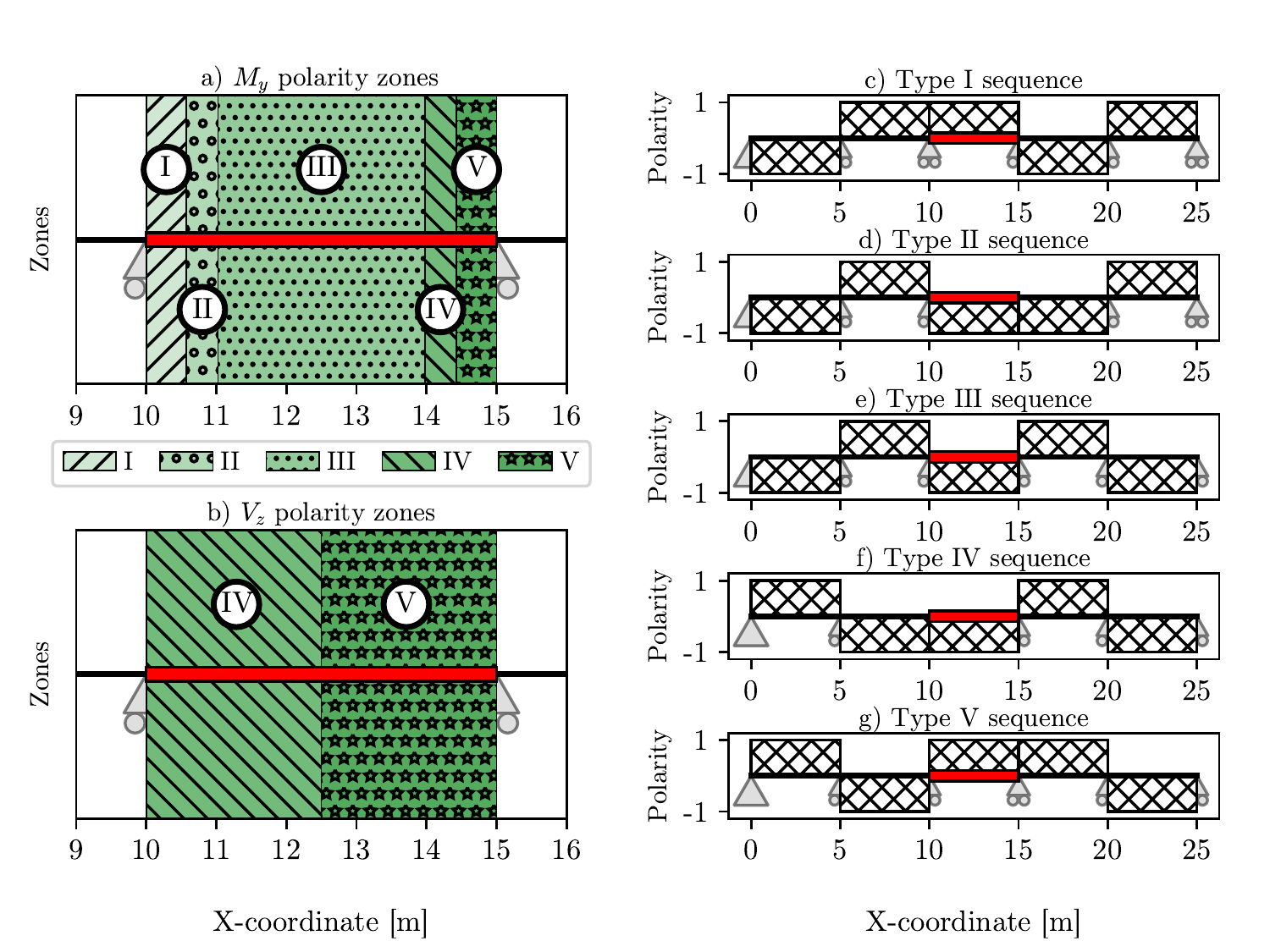}
	\caption{Polarity zones that occur along various span segments of a $m=5$ homogeneous beam system of equal span and cross-sectional properties. The same zones and sequences, although at different boundaries, occur in heterogeneous (varying UDL and span) systems.}
	\label{fig:polarity_zones_and_patterns}
\end{figure*}

A rigorous qualitative study of the polarity sequences for different IL locations and design scenarios revealed 5 unique polarity sequences that occur along specific segments of a given beam span termed \textit{polarity zones}, which are illustrated in Figure \ref{fig:polarity_zones_and_patterns} for the central beam highlighted in red.

These 5 polarity zones are common to all members of both homogeneous (equal spans and cross-sections) as well as heterogeneous continuous beam systems, although the exact boundaries between one zone varied depending on the relative magnitude of spans and cross-section properties. The sequences identified in Figure \ref{fig:polarity_zones_and_patterns} also apply to larger beam systems with the polarity direction alternating at each successive beam. For example, if the 5-member system was extended by an additional member on either side of the system (to give a 7 member system), the left-most member of the Type I polarity sequence would have a positive polarity, and similarly, the right-most member would have a negative polarity. The same logic extends to the other four sequences.

Each polarity sequence is indicative of two critical load arrangements that maximise the positive or negative internal member forces respectively. The maximum positive load arrangement for Type I is also equal to the maximum negative load arrangement for Type IV, since these sequences are polar opposites of each other, which is also true for the Type II and Type IV sequences. Consequently, these 5 polarity zones correspond to 6 unique critical load arrangements for a given beam, namely positive Type I, II and III along with their (negative) polar opposites. The only exceptions occur for the beams at either end of the spans, named end-span beams, in which the Type I and Type IV sequences collapse into the Type III sequence (or its polar opposite) at the left end, and similarly for the Type II and Type V sequences at the right end, resulting in four unique load arrangements for end-span beams.

\begin{figure*}[htb]
	\centering
    \includegraphics[width=\textwidth,height=\textheight,keepaspectratio]{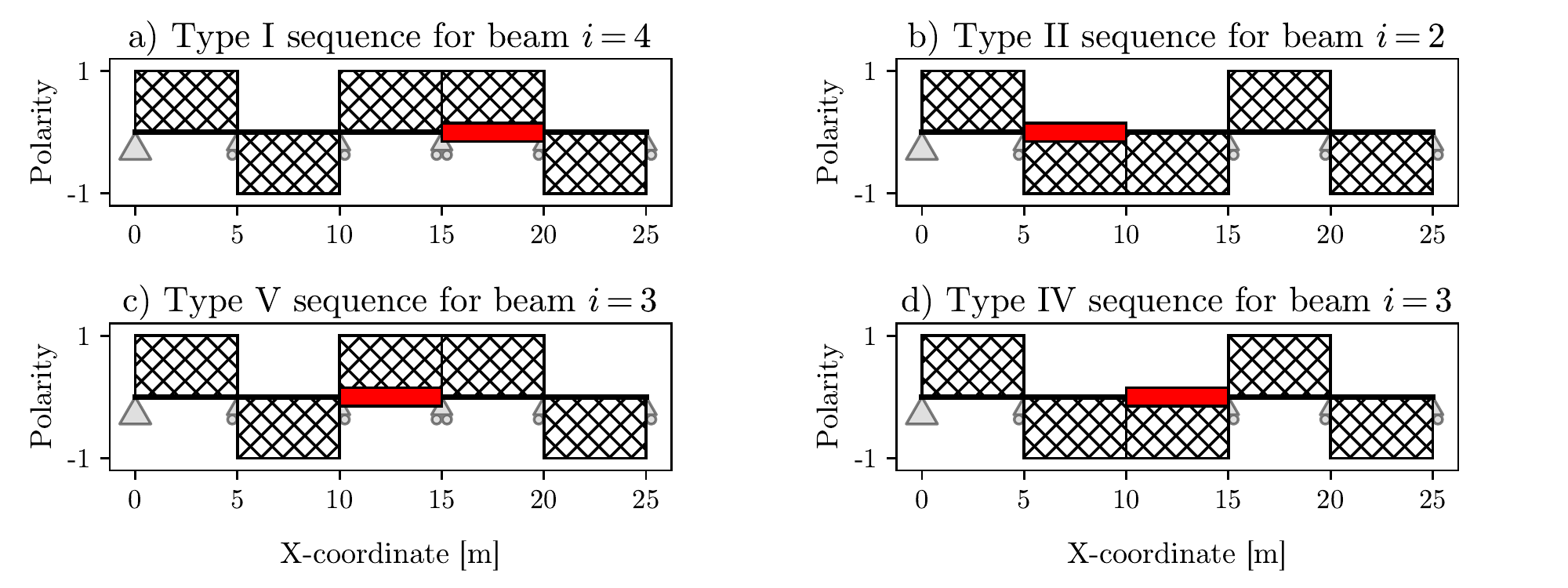}
	\caption{Polarity sequences are identical for adjacently lying beams (highlighted in red) for Type I and Type V sequences as shown by Figure a) and Figure c), as well as Type II and Type IV sequences, as shown by Figure b) and Figure d).}
	\label{fig:polarity_sequence_simplification}
\end{figure*}

\begin{figure*}[htb]
	\centering
    \includegraphics[width=\textwidth,height=\textheight,keepaspectratio]{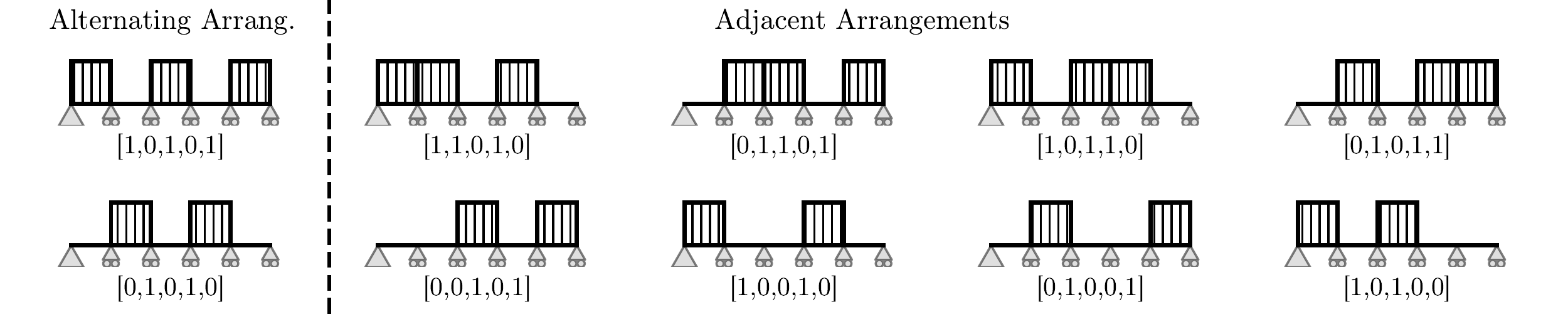}
	\caption{The critical load arrangements set $\mathbf{J_\mathrm{flex}}$ of size $p=2m$ for a 5-member continuous beam system ($p=10$) grouped in alternating and adjacently loaded arrangements.}
	\label{fig:critical_load_arrangements_2m_5members}
\end{figure*}

Whilst each non-end-span beam has 6 unique critical load arrangements, it does not mean that the beam system has $6m$ unique load arrangements ($m$ is the number of members in the beam system). This is because, as shown in Figure \ref{fig:polarity_sequence_simplification}, the maximum positive Type V load arrangement for one beam is identical to the maximum positive Type I load arrangement of the beam immediately adjacent to (the right of) it. A similar overlap exists between Type II and Type IV sequences, and the two Type III load arrangements (for maximum and negative internal forces) are identical for all beams. Through a process of elimination, it is possible to simplify the actual total number of potential critical load arrangements to $p_\mathrm{flex} = 2m$. This set will be termed the \textit{flexural load arrangements} set $\mathbf{J_\mathrm{flex}}$, and can be evaluated using Algorithm \ref{algo:flexural_set} provided in \ref{app:algorithm_1}, with an example output for a $m=5$ system shown in Figure \ref{fig:critical_load_arrangements_2m_5members}, grouped in alternating and adjacently loaded arrangements.

The load arrangement set $\mathbf{J_\mathrm{flex}}$ of size $p_\mathrm{flex} = 2m$ identified here is a literal exponential improvement to the brute-force approach of analysing and designing against $p = 2^m$ load arrangements and for evaluating the influence zone with Equation \ref{eq:global_formulation}. It needs to be shown though that all critical load arrangements $\mathbf{J_\mathrm{crit}}$ fall within $\mathbf{J_\mathrm{flex}}$ (i.e. $\mathbf{J_\mathrm{crit}} \in \mathbf{J_\mathrm{flex}}$).

\subsection{Shear beams and the impact on critical load arrangements}

To check if the load arrangement set $\mathbf{J_\mathrm{flex}}$ contains all critical pattern loads, various continuous beam systems of size up to $m=10$ (to facilitate computational feasibility) were numerically generated, using randomly distributed UDLs and span lengths based on the design constraints identified in Table \ref{table:design_sets}. By calculating all $2^m$ load arrangements, evaluating the utilisation ratio (based on the moment and shear force combinations) of each, the load arrangement that caused the worst-case utilisation ratio could be identified. This set of critical load arrangements was then compared against the $\mathbf{J_\mathrm{flex}}$ set identified by Algorithm \ref{algo:flexural_set} provided in \ref{app:algorithm_1}.

Although $\mathbf{J_\mathrm{flex}}$ tended to cause the critical utilisation ratio in the majority of cases, there were instances were other load arrangements not within $\mathbf{J_\mathrm{flex}}$ controlled the design. This unexpected behaviour occurred in cases where short spanning, deep beams were included. Analysing these special cases in detail indicated that the exact conditions under which the previously unidentified load arrangements occur were generally related to the following $L_\mathrm{shear}$ span limit quantified by

\begin{equation}
    \label{eq:shear_beam_span}
    \sqrt{\frac{6 E I_{yy}}{G A_z}} < L_\mathrm{shear}
\end{equation}

\noindent where $E$ and $G$ are the Young's and shear modulus of the material respectively, and $I_{yy}$ and $A_z$ are the major second moment of area and shear area of the prismatic beam, respectively.

\begin{figure*}[!htb]
	\centering
    \includegraphics[width=\textwidth,height=\textheight,keepaspectratio]{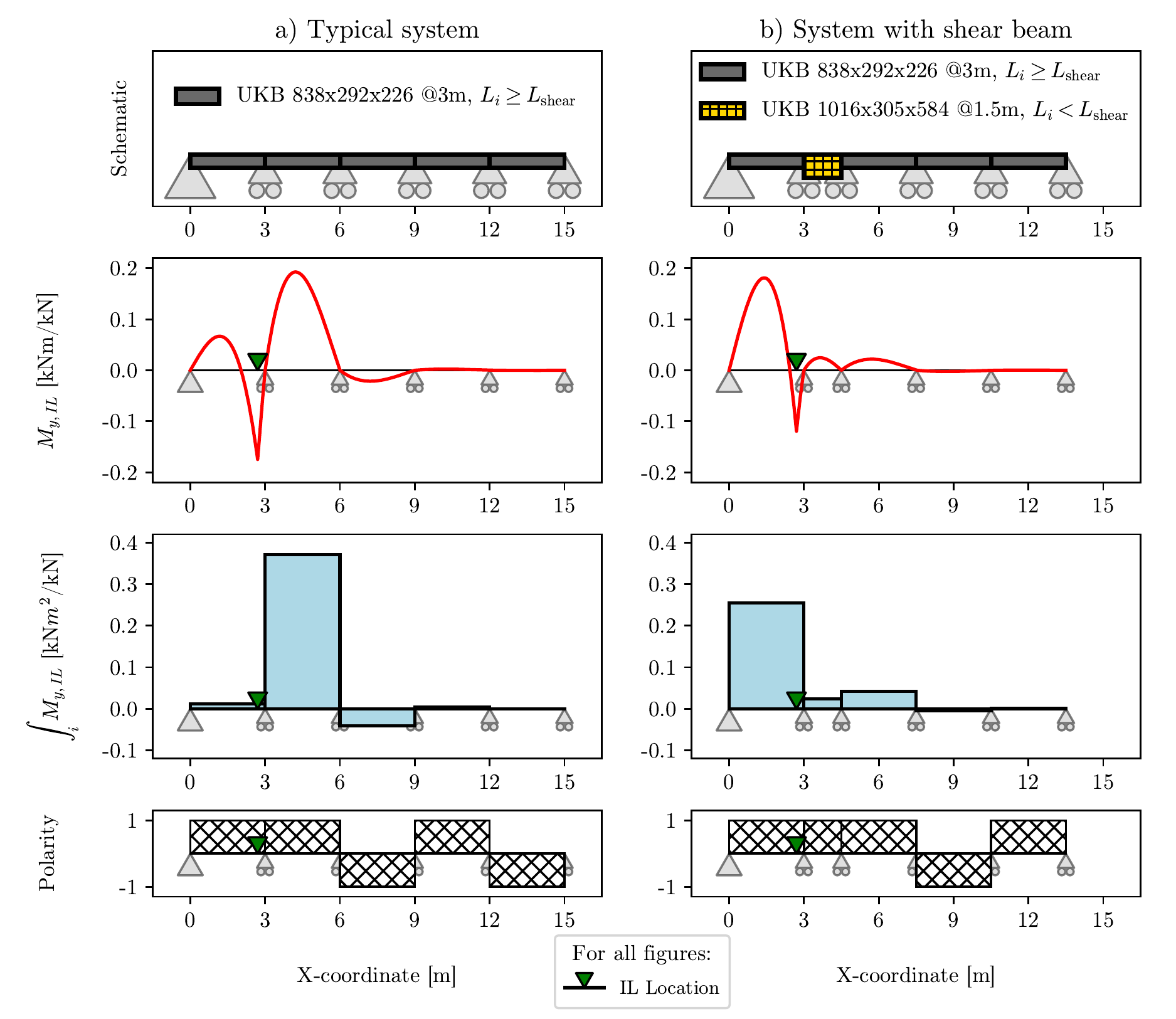}
	\caption{A schematic demonstrating the impact of a shear beam (highlighted in yellow) on a standard polarity sequence of a continuous beam system when spans shorter than the shear span limit $L_\mathrm{shear}$ (as identified by Equation \ref{eq:shear_beam_span}) occur. Note the flipped polarity directions of the members on the right-hand side of the system.}
	\label{fig:shear_beam_impact}
\end{figure*}

Although the $L_\mathrm{shear}$ span limit appears to be related to shear beams, this is the first time that shear beams have been reported in literature to cause novel critical load arrangements. As shown in Figure \ref{fig:shear_beam_impact}, shear beams appear to flip the polarity of the immediately adjacent member when measured outwardly from a given IL location , with all subsequent members alternating the polarity direction as before.

When shear beams (as defined by the $L_\mathrm{shear}$ limit) occur, they introduce new critical load arrangements not found within $\mathbf{J_\mathrm{flex}}$. The increase in terms of the final utilisation factor of the beams was typically in the range of 4-5\%, although larger increases were also observed. Whilst a thorough analysis of the increase in utilisation ratio caused by these newly identified load arrangements would be of interest, it falls outside the scope of this study. Instead, an algorithm will be presented capable of identifying these new load arrangements \textit{a priori}, which is the main objective of this investigation as explained in section \ref{sec:key_challenge}.

\subsection{Determining shear beam induced critical load arrangements \textit{a priori}}
\label{sec:shear_load_arrangements}
The principal issue when evaluating the shear beam induced critical load arrangements \textit{a priori}, hereafter referred to as the \textit{shear load arrangements} $\mathbf{J_\mathrm{shear}}$, is the fact that the final material and cross-sectional properties to evaluate the $L_\mathrm{shear}$ limit in Equation \ref{eq:shear_beam_span} are not known until the beam is designed. This creates a causality dilemma and hence needs to be addressed.

In clear opposition to the $\mathbf{J}_\mathrm{flex}$ set, which does not depend on the continuous beam system properties, the shear load arrangements cannot be established \textit{in universum} without some system knowledge. However, by taking advantage of the design constraints set by Equation \ref{eq:design_constraints}, one can identify \textit{a priori} what members are potentially susceptible to cause shear load arrangements by re-writing Equation \ref{eq:shear_beam_span} as:

\begin{equation}
    \label{eq:max_shear_beam_span}
    \sqrt{6 \left( \frac{E}{G} \right)_\mathrm{max}  \left( \frac{I_{yy}}{A_z} \right)_\mathrm{max} } < L_\mathrm{shear, max}
\end{equation}

The above equation groups the maximum material and cross-sectional property ratios together. 
By limited the design space to S355 steel and UKB section sizes as specified in Section \ref{sec:design_constraints}, the maximum material ratio ($(E/G)_\mathrm{max} = 2.600$) and cross-sectional property ratio ($(I_{yy}/A_{z})_\mathrm{max} = 0.397 m^2$) can be evaluated. Consequently, beams shorter than the shear span limit 
are susceptible to cause shear load arrangements (in this case $L_\mathrm{shear, max} = 2.49$m). In identifying these susceptible members \textit{a priori}, it is possible to evaluate the shear load arrangements using Algorithm 2 provided in \ref{app:algorithm_2}.

Algorithm 2 transforms the flexural load arrangement from set $\mathbf{J_\mathrm{flex}}$ based on a list of susceptible shear beams identified by Equation \ref{eq:max_shear_beam_span}. This is achieved by flipping the on/off activation factor of the load arrangement if a shear beam is encountered whilst travelling outwardly in both the left (-1) and right (1) direction from a start beam index. This operation transforms the flexural load arrangement based on the behaviour identified visually in Figure \ref{fig:shear_beam_impact}, and needs to check four individual case conditions to account for continuous beam systems that have multiple, potentially adjacently lying, shear beams.

Since every beam system is of size \textit{m}, the time complexity of a single pass of Algorithm \ref{algo:shear_set} is $O(m)$. However, since every flexural load arrangement ($2m$), and every combination of $n$ potential shear beams ($2^n-1$ combinations, as the zero set is already considered in $\mathbf{J_\mathrm{flex}}$ by default), and every possible start-index ($m$) needs to be computed, the time complexity to evaluate the shear set $\mathbf{J_\mathrm{shear}}$ would be $O(m^3\, 2^n)$. It should be noted that this process is computationally expensive.

It was observed that passing every possible start index generated either duplicate shear load arrangements, or occasionally existing flexural load arrangements. For example, for a given singular potential shear beam location, the algorithm would result in the same transformed shear load arrangement for all start-indices starting on the left and right hand-side of that susceptible shear beam location. Similarly, the two alternating arrangements from $\mathbf{J_\mathrm{flex}}$ would result in an already existing adjacent arrangement from $\mathbf{J_\mathrm{flex}}$ if only a singular susceptible shear beam exists.

Using such logic, it is sufficient to pass only adjacent arrangements from $\mathbf{J_\mathrm{flex}}$ along with the left-hand (or right-hand) index of the adjacently loaded spans as the start index for Algorithm \ref{algo:shear_set} to yield an effective set of potential shear load arrangements. By not having to evaluate Algorithm \ref{algo:shear_set} for every possible start index of each load arrangement, the computational complexity reduces to $O(m^2\, 2^n)$. From this, it also follows that since the alternating load arrangement is never transformed  (which leaves only $2(m-1)$ load arrangements to be passed to the algorithm) and since $2^n-1$ possible shear beam combinations can exist, the maximum number of unique critical shear load arrangements should be of size $p_\mathrm{shear} = 2(m-1)(2^n-1)$.

\subsection{Validating flexural and shear load arrangement algorithms}
\label{sec:validation_algos}
A design data set consisting of $32$ UDL and $32$ span values sampled from a random uniform distribution for a $m=10$ beam system was generated based on the high-variation design scenario identified in Section \ref{sec:design_constraints}. Significantly higher variable UDLs ($Q_{k,i} \in [200 \text{\,kN/m}, 400 \text{\,kN/m} ]$) were applied to increase the likelihood of deep beams and thereby critical shear load arrangements, allowing the performance of the algorithm to be stress-tested. This resulted in $10\times 32\times 32 = 10240$ individual beam design examples, for which the critical load arrangement $J_\mathrm{crit}$ could be identified.

\begin{figure*}[!htb]
	\centering
    \includegraphics[width=\textwidth,height=\textheight,keepaspectratio]{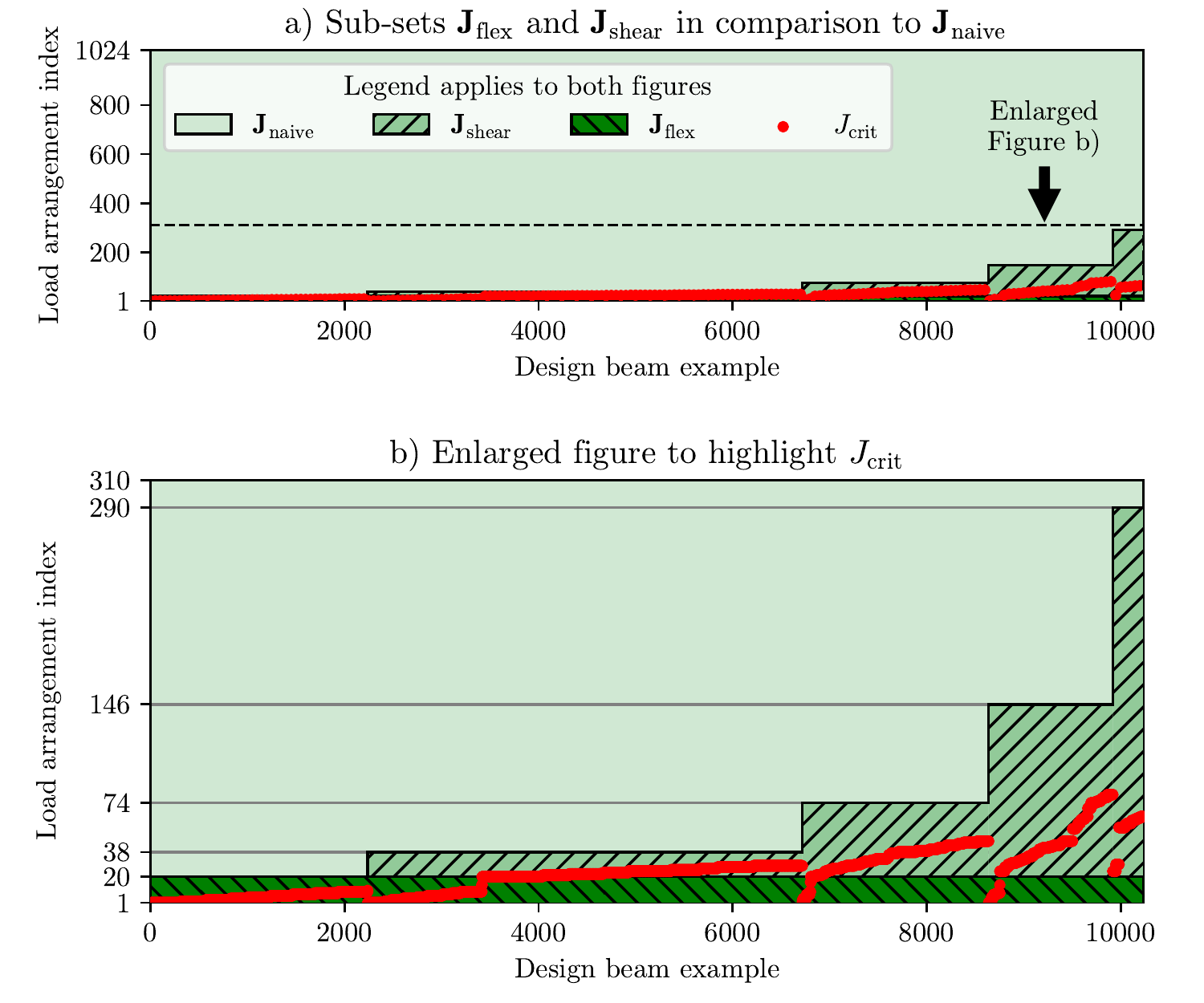}
	\caption{Load arrangement index for each design beam example ordered in increasing number of shear beam occurrences and critical load arrangement indices. This confirms visually that the critical load arrangement $J_\mathrm{crit}$ for each design beam example from the generated data set falls within either $\mathbf{J_\mathrm{flex}}$ or $\mathbf{J_\mathrm{shear}}$ and are significantly smaller than $\mathbf{J_\mathrm{naive}}$. Figure b) is an enlarged view of Figure a).}
	\label{fig:validation_algorithms}
\end{figure*}

The results of this validation exercise are illustrated in Figure \ref{fig:validation_algorithms}, which plots the critical load arrangement index for each design beam example. Every load arrangement index corresponds to a unique load arrangement out of the naive set $\mathbf{J}_\mathrm{naive}$ of size $p_\mathrm{naive}=2^m=1024$. The set $\mathbf{J}_\mathrm{naive}$ was ordered so that the load arrangements for set $\mathbf{J}_\mathrm{flex}$ are first, followed by those of set $\mathbf{J}_\mathrm{shear}$, and subsequently all others. The design examples themselves were sorted twice: first in ascending number of shear beam occurrences, and subsequently in ascending load arrangement indices. This results in the gradual increase of the $J_\mathrm{crit}$ indices as seen in Figure \ref{fig:validation_algorithms}.

Figure \ref{fig:validation_algorithms} sheds insight on a number of important points. The first is that the critical load arrangement $J_\mathrm{crit}$ for every single beam example from the 10240 data set occurred within the $\mathbf{J_\mathrm{flex}}$ or $\mathbf{J_\mathrm{shear}}$ sets, validating the qualitative analysis based on the polarity zones and sequences identified previously. This also emphasises the validity of Algorithm \ref{algo:flexural_set} and \ref{algo:shear_set}. Furthermore, the set size predictions $p_{flex} = 2m$ and $p_{shear} = 2(m-1)(2^n-1)$ are also confirmed. For the $m=10$ member system designed here, $p_{flex} = 20$, and depending on the number of shear beam occurrences of each system, which varied from $n=\{0, 1, 2, 3, 4\}$, the number of shear load arrangements varied from $p_{shear}=\{0, 18, 54, 126, 270\}$. This corresponded to $p_{total}=\{20, 38, 74, 146, 290\}$ respectively, as indicated by the $y$-axis of Figure \ref{fig:validation_algorithms} b).

Figure \ref{fig:validation_algorithms} a) also emphasises how much smaller sets $\mathbf{J_\mathrm{flex}}$ and $\mathbf{J_\mathrm{shear}}$ are in comparison to $\mathbf{J_\mathrm{naive}}$. This will greatly reduce the number of load-arrangements that need to be analysed, reducing the computational cost of both optimally designing the continuous beam system and evaluating the influence zone for system lengths of $m>10$. Further insights generated by Figure \ref{fig:validation_algorithms} are discussed in Section \ref{sec:discussion}.

\subsection{Summary of critical load arrangements of continuous beams}
\label{sec:summary_J_crit}
By adding the set of critical flexural and shear load arrangements together, it is possible to explicitly define \textit{a priori} the set of critical load arrangements for any continuous beam system under defined design constraints. For the purpose of the influence zone concept and Equation \ref{eq:global_formulation}, it is assumed that: $\, \mathbf{J} \rightarrow \mathbf{J_\mathrm{crit}} \in \mathbf{J_\mathrm{flex}} \cup \mathbf{J_\mathrm{shear}}$. The results from this systematic critical load arrangement investigation are summarised in Table \ref{table:crit_patterns_sets}.

\begin{table*}[!htb]
	\centering
    \small
	\begin{tabularx}{0.8\textwidth}{c cc}
		\toprule
		\multicolumn{1}{c}{Set} & Set Size & Algorithm Complexity \\ \midrule
		\multicolumn{1}{>{\centering\arraybackslash}X}{Critical load arrangements per internal beam} & 6 & $O(1)$ \\ [.8cm]
  		\multicolumn{1}{>{\centering\arraybackslash}X}{Critical load arrangements per end-span beam} & 4 & $O(1)$ \\ [.8cm]
		\multicolumn{1}{>{\centering\arraybackslash}X}{$\mathbf{J_\mathrm{flex}}$ - Critical flexural arrangements per beam system } & $2m$ & $O(m)$ \\ [.8cm]
		\multicolumn{1}{>{\centering\arraybackslash}X}{$\mathbf{J_\mathrm{shear}}$ - Critical shear arrangements per beam system } & $2(m-1)(2^n-1)$ & $O(m^2\, 2^n)$ \\ [.8cm]
		\multicolumn{1}{>{\centering\arraybackslash}X}{$\mathbf{J_\mathrm{naive}}$ - Naive load arrangements } & $2^m$ & $O(2^m)$ \\
		\bottomrule
	\end{tabularx}
  	\caption{Load arrangements set summary for $m$ dimensional beam systems containing $n$ shear beams with associated algorithm complexities.}
	\label{table:crit_patterns_sets}
\end{table*}

%% file: 04_Influence_Zones.tex
\section{Influence zone evaluation}
\label{sec:influence_zones}

\subsection{Explicitly defining the utilisation ratio contribution function}

By taking advantage of the concept of integrated influence lines and polarity zones from Section \ref{sec:polarity_diagrams}, and by having explicitly defined the critical load arrangement set $\mathbf{J} \rightarrow \mathbf{J_\mathrm{crit}}$ in Section \ref{sec:summary_J_crit}, it is possible to define the utilisation ratio contribution function $\mathbf{u}_{d,i,\,j}$ as:

\begin{equation}
    \begin{gathered}
    \label{eq:utilization_d_i_j}
    \mathbf{u}_{d,i,\,j} \rightarrow \mathbf{D_\mathrm{ULS}}(I_d, M_{d, i,j}, V_{d, i,j}) \\
        M_{d, i,j} = w_i \ J_{i,j} \int_{i} \mathbf{M}_{\mathrm{IL}, d} \\
        V_{d, i,j} = w_i \ J_{i,j} \int_{i} \mathbf{V}_{\mathrm{IL}, d} \\
    \end{gathered}
\end{equation}

\noindent $\mathbf{D_\mathrm{ULS}}$ represents the ULS steel cross-section design checks based on Eurocode EN 1993-1-1 6.2 \cite{CEN_2015_BS_EN_1993_1_1_2005_A1_2014}, $I_d$ represents the cross-sectional properties, $M_{d,i,j}$ denotes the major axis moment while $V_{d,i,j}$ is the major axis shear force of the design beam $d$, $w_i$ is the UDL, and $J_{i,j}$ is the activation factor of the load arrangement $j$ from the set $\mathbf{J_\mathrm{crit}}$ for beam $i$. Integrals $\int_{i} \mathbf{M_\mathrm{IL, d}}$ and $\int_{i} \mathbf{V_\mathrm{IL, d}}$ are the integrated influence line values across beam $i$ for a particular influence line location within the design beam $d$ as introduced in Figure \ref{fig:influence_lines_to_polarity_zones}.

The influence line locations within the design beam need to correspond with the worst-case internal force locations. Whilst engineering experience would dictate those to occur over the supports, they can in fact arise anywhere along the design beam depending on the exact distribution of UDLs, spans and cross-sectional properties, since each segment of the design beam has its own critical load arrangement as highlighted by Figure \ref{fig:polarity_zones_and_patterns}. In this study, a total of 11 influence line locations were sampled, one at either support and another 9 equidistantly distributed between the supports.

For a given design beam and $k_\mathrm{max}$ value, Equation \ref{eq:utilization_d_i_j} will therefore result in $11p$ utilisation ratios (recall that $p$ is the set size of $\mathbf{J}$), from which the critical utilisation ratio (the maximum one) is evaluated in Equation \ref{eq:global_formulation} to check if the $\epsilon_\mathrm{max}$ threshold has been attained. Note that as $k_\mathrm{max}$ increases, the critical influence line location and the critical load arrangement can vary, yet as $k_\mathrm{max}$ approaches $m$, they will equate to the location and load arrangement that governed the design for that particular beam.

\subsection{Design data set generation}
The size of the continuous beam system $m$ to be modelled needs to be at least double the maximum influence size $k_\mathrm{max}$. This is because the highest influence zone measurable for the middle span of a continuous beam is by design half the system length $m$. Therefore, size $m$ needs to be chosen such that $\max(\mathbf{k_\mathrm{max}}) < m/2$, where $\mathbf{k_\mathrm{max}}$ is the list of all influence values $k_\mathrm{max}$ of the continuous beam system. Since evaluating $k_\mathrm{max}$ is the main aim of this investigation, a sufficiently large value for $m$ needs to be assumed; $m = 15$ was used for this purpose. 

Individual design data sets consisting of $32$ UDL and $32$ span values sampled from a random uniform distribution for a $m=15$ beam system were created based on the design constraints identified in Section \ref{sec:design_constraints} for Sets 2, 3, and 4, each containing $32 \times 32 \times 15 = 15360$ beam designs. For Set 1, the difference within the beam systems only varied in terms of the identical span $L$ and UDLs $Q_k$ of the beams, which were also sampled in $0.5 \text{\ m}$ and $5 \text{\ kN/m}$ increments respectively. Given that this results in 23 span and 13 UDL increments for Set 1 respectively, Set 1 contained $23 \times 13 \times 15 = 4485$ design examples.

For the design optimisation of the continuous beam systems, a coupled analysis and design approach was taken, optimising for minimum structural depth. Design sensitivity analysis was avoided by an implicit ordering of the UKB section list based on structural capacity. The influence zone values were extracted using Equation \ref{eq:global_formulation} and Equation \ref{eq:utilization_d_i_j}.

\subsection{Influence zone results}

\begin{figure*}[!b]
	\centering
    \includegraphics[width=\textwidth,height=\textheight,keepaspectratio]{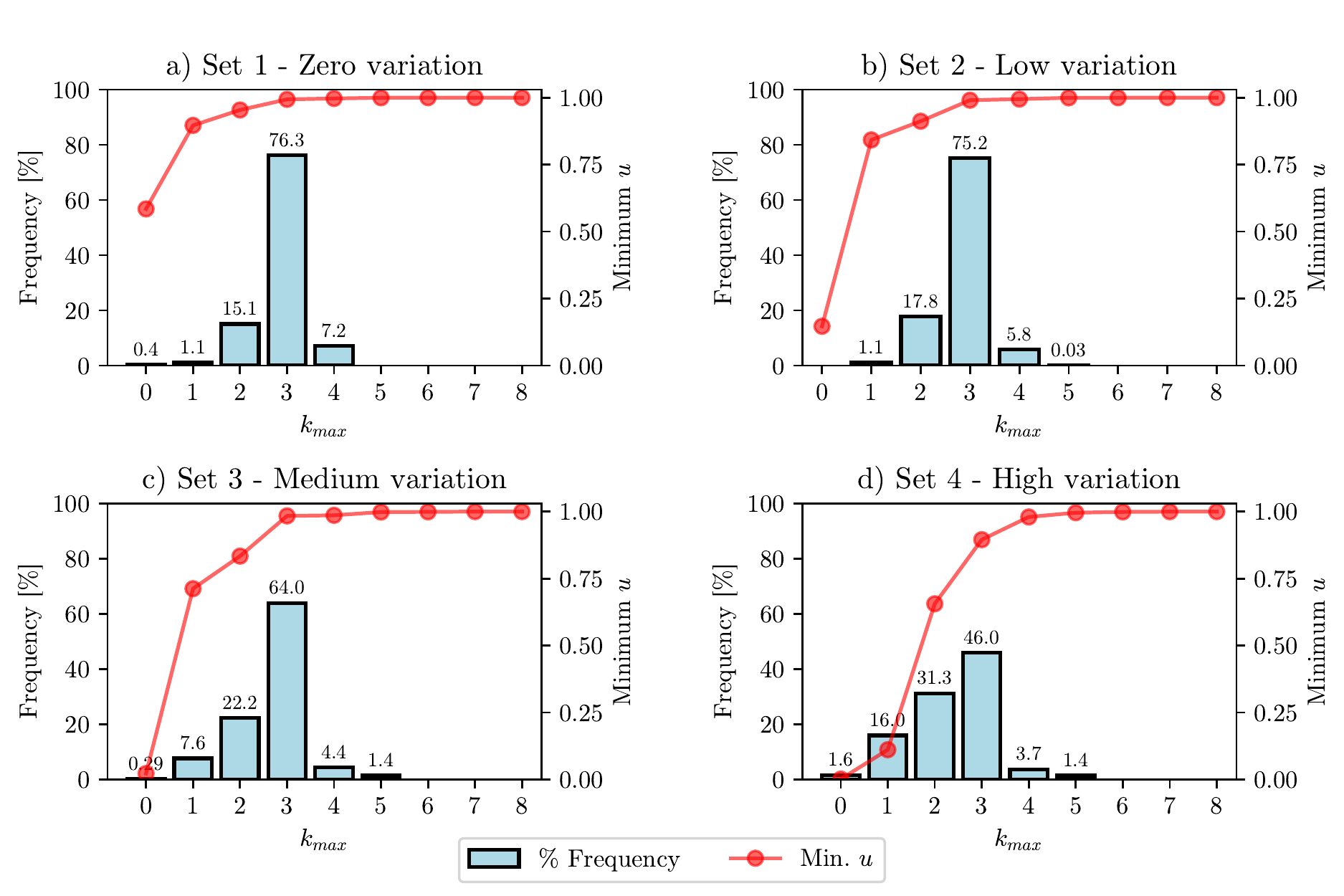}
	\caption{Influence zone results for various design constraints with a max error threshold $\epsilon_\mathrm{max} = 0.005$ indicating the percentage frequency distributions of the influence zone values $k_\mathrm{max}$ and minimum utilisation factors captured for each $k_\mathrm{max}$ value for a given design beam $d$.}
	\label{fig:influence_zone_results}
\end{figure*}

The influence zone results are shown in Figure \ref{fig:influence_zone_results} for a max error threshold $\epsilon_\mathrm{max} = 0.005$ and various design data sets defined in Section \ref{sec:design_constraints}. For all sets investigated, the most common influence zone value (the mode) was $k_\mathrm{max} = 3$, and the majority of influence zone values were at $k_\mathrm{max} \le 3$, meaning the span and applied loading information of a given beam along with that of the three adjacent spans on either side captured the correct utilisation ratio of the design beam with less than a $\pm 0.5\%$ error in the majority of cases.

However, the various sets reveal differences in the maximum and distribution of the influence zone. The maximum influence zone value for Set 1 was $k_\mathrm{max}=4$, whereas it was $k_\mathrm{max}=5$ for Set 2, 3 and 4. Furthermore, as the set number increases, which corresponds with an increase in variation of the design information in terms of spans and UDLs, the influence zone value distribution appears to flatten and widen. For example, it was the high-variation Set 4 which actually contained the most influence zone values $k_\mathrm{max}=0$ for 1.6\% of the design examples, whereas the zero variation Set 1 only had 0.4\% of its design examples exhibit an influence zone of $k_\mathrm{max}=0$. The minimum utilisation curve (red curve with point markers in Figure \ref{fig:influence_zone_results}) captured by each influence zone value suggests that, in general, increasing design variation leads to greater maximum influence zone values.

The average and maximum influence zone values were also calculated for various error thresholds as shown in Table \ref{table:influence_zone_results}. Note that the maximum influence zone value of $k_\mathrm{max}=7$ for Set 4 with the highest error threshold confirms that the $m=15$ member-size assumption was sufficient for the purpose of this study. Together Figure \ref{fig:influence_zone_results} and Table \ref{table:influence_zone_results} provide evidence for the following conclusions:

\begin{itemize}
    \item A decrease in the acceptable error threshold correlates with an increase in both the average and maximum influence zone range.
    \item An increase in design variation correlates with an increase in the maximum influence zone range.
    \item An increase in design variation, however, correlates with a decrease in average influence zone range in most instances where the acceptable error threshold is relatively tight ($\epsilon_\mathrm{max} \le 10\%$). At higher error thresholds the trend is less discernible.
\end{itemize}

\begin{table*}[!ht]
    \centering
    \small
    \begin{tabularx}{0.82\textwidth}{cccccccccc}
        \toprule
        \multirow[b]{2}{1.5cm}[-0.05cm]{\centering Error $\epsilon_\mathrm{max}$ [\%]} &
        \multicolumn{4}{c}{Average $k_\mathrm{max}$} &&
        \multicolumn{4}{c}{Maximum $k_\mathrm{max}$} \\ \cmidrule{2-5} \cmidrule{7-10}

        & Set 1 &  Set 2 & Set 3 & Set 4 && Set 1 & Set 2 & Set 3 & Set 4 \\ \midrule
        
        0.1 & 4.60 & 4.46 & 3.84 & 3.37 && 5 & 5 & 6 & 7 \\
        0.5 & 2.89 & 2.86 & 2.69 & 2.38 && 4 & 5 & 5 & 5 \\
        1 & 2.76 & 2.75 & 2.35 & 2.06 && 3 & 3 & 5 & 5 \\
        5 & 1.52 & 1.39 & 1.29 & 1.17 && 2 & 3 & 3 & 4 \\ 
        10 & 0.98 & 0.98 & 0.89 & 0.83 && 2 & 2 & 3 & 4 \\
        20 & 0.76 & 0.84 & 0.73 & 0.67 && 1 & 1 & 2 & 3 \\
        50 & 0.00 & 0.30 & 0.41 & 0.43 && 0 & 1 & 1 & 2 \\ \bottomrule
    \end{tabularx}
    \caption{Influence zone results for various maximum error thresholds $\epsilon_\mathrm{max}$ for each design data set, evaluating average and maximum influence zone values $k_\mathrm{max}$. Note that increasing set numbers corresponds with increasing design variation, a proxy for design complexity, see Table \ref{table:design_sets} for details.}
	\label{table:influence_zone_results}
\end{table*}

It should be noted that an error threshold of less than $0.5\%$ is relatively small in comparison to uncertainties that exist in structural design.
These uncertainties include, for example, material yield strength and imposed UDL values (consider that variable UDL values $Q_k$ are increased $50\%$ with a load combination factor of $1.5$ within the Eurocodes \cite{CEN_2010_BS_EN_1990_2002_A1_2005}).
Furthermore, the design constraints of design set 4 represent the top end of design variation which may occur in typical continuous beam systems.
Consequentially, it is reasonable to suggest that for continuous beam systems with design constraints specified in section \ref{sec:design_constraints} the influence zone values are on average $k_\mathrm{max}<3$, and in the most extreme case $k_\mathrm{max} = 5$.

%% file: 05_Discussion.tex
\section{Discussion}
\label{sec:discussion}

The results along with the methodology of the influence zone investigation has led to a number of important findings. These include introducing the novel concept of the \textit{structural influence zone} and a numerical methodology of evaluating it, discovering novel \textit{shear load arrangements} with the help of \textit{polarity zones} and \textit{polarity sequences}, and introducing \textit{load arrangement algorithms} to explicitly identify critical load arrangements of continuous beam systems of any arbitrary member size, which were a necessary prerequisite for the influence zone study. Each of these findings are discussed in detail and contextualised with relevant existing literature.

\subsection{Influence zone insights}

The influence zone results confirm that the impact of loading, and by extension of any design information, drops off sharply the further away one moves from the influence line location. This behaviour can be identified across all influence line diagrams found within this paper, such as Figures \ref{fig:influence_lines_to_polarity_zones} and \ref{fig:shear_beam_impact}. This investigation has formulated this concept as the \textit{influence zone}, shown how it applied to continuous beam systems, and rigorously studied the influence zone distributions under various design assumptions and error thresholds. 

An important element of the influence zone definition in Section \ref{sec:mathematical_formulation} should be brought to light. The influence zone value $k_\mathrm{max}$ is only found when two important conditions are met, notably when both the smallest value of $k_\mathrm{max}$ and all values larger than this value of $k_\mathrm{max}$ conform to Equation \ref{eq:global_formulation}. This was a necessary constraint to account for the fact that the ratio $u_{d,cap} / u_{d,true}$ sometimes converged towards unity by oscillation.

The cause of this behaviour results from adjacent load arrangements (see Figure \ref{fig:critical_load_arrangements_2m_5members}) which were sometimes more critical when only a segment of (as opposed to the entire) load arrangement was considered. For example, the maximum $u_{d,cap} / u_{d,true}$ ratio calculated for Set 4 when assuming an influence zone value of $k_\mathrm{max} = 1$ was 1.89, only for the ratio to drop to 0.942 and 0.998 for influence zone values $k_\mathrm{max} = 2$ and $k_\mathrm{max} = 3$, respectively. Future research on influence zones should keep this non-intuitive behaviour in mind, since a simple $u_{d,cap} / u_{d,true} < r_{cap}$ threshold, in which $r_{cap}$ represents a minimum threshold of captured utilisation would lead to an underestimation of the influence zone distribution.

\subsection{Demarcating influence zones from influence lines}

Although there is a proximal relationship between the concept of influence zones and influence lines, mostly evidenced by Equation \ref{eq:utilization_d_i_j} where integrated influence lines play an important role for the evaluation of influence zones, these two concepts differentiate themselves in important ways. This distinction also applies to the two-dimensional application of influence lines known as influence surfaces \cite{deng_genuine_2023, zheng_bridge_2021, orakdogen_direct_2005, memari_computation_1991}.

Whilst influence lines/surfaces are exact analytical tools that define the mechanical response of a known structural system about a particular point, influence zones are a heuristic design tool that offer insight on what information is relevant to the design of the structural system to begin with based on certain analytical assumptions. The value of influence lines/surfaces arise during analysis on a system-by-system basis, whereas the value of influence zones arise during design after having studied them in their statistical aggregate.

This distinction could be considered further evidence supporting the demarcation between design and analysis in structural engineering. Previous literature has highlighted the difference between \textit{knowledge-that} explains fundamental facts about systems (such as influence lines) versus \textit{knowledge-how} something can be designed or solved (such as influence zones) \cite{bulleit_what_2012, bulleit_philosophy_2015}. Recent literature has suggested that the processes of analysis and design solve related, albeit oppositely posed problems known as forward and inverse problems respectively \cite{gallet_structural_2022}. Influence lines can be seen as a tool that solves the former, whereas influence zones solve the latter.

As a matter of fact, the influence zone concept was developed whilst developing a design model for continuous beam systems from an inverse problem perspective, and allows the \textit{a priori} knowledge of what span and loading information is relevant for design of a particular continuous beam. It is possible that the influence zone concept could serve as an important heuristic tool in the design of continuous structural systems, supporting the view that the application of heuristics is a cornerstone for engineering design \cite{koen_discussion_2003}. Further novel ideas might be uncovered when approaching engineering design from an inverse problem perspective.

\subsection{Flexural load arrangements}
An important contribution of this investigation was presenting the flexural load arrangements clearly through the use of \textit{polarity sequences}. Notably the \textit{polarity zones} highlight which load arrangement is critical for specific segments of a beam, which could be useful in the design of tapered (non-prismatic) continuous beam systems \cite{kaveh_optimal_2018, veenendaal_design_2011}.

The influence zone study allows the contextualisation of simplified load arrangement provisions. For example, whilst Annex AB.2 from EN 1993-1-1 \cite{CEN_2015_BS_EN_1993_1_1_2005_A1_2014} covers alternating flexural load arrangements in full, it specifies that for the adjacent flexural load arrangement type, the two adjacently loaded spans are the only spans required to factor the variable load ($Q_k$). In essence, the variable load information on all other spans aside from the beam under consideration and the two directly adjacent spans are ignored, which is the technical equivalent of assuming an influence zone to $k_\mathrm{max}=1$.

With help of Table \ref{table:influence_zone_results}, it is possible to infer that an influence zone value $k_\mathrm{max}=1$ is likely to introduce an error between $5-10\%$ in terms of the true utilisation for design scenarios with no UDL or span variation (the average $k_\mathrm{max}$ value for $\epsilon_\mathrm{max} = 5\%$ and $\epsilon_\mathrm{max} = 10\%$ is 1.52 and 0.98 for Set 1 respectively). The simplified Eurocode provisions are therefore, on average, a reasonable simplification to capture the impact of variable load arrangements. However, the maximum influence zone value of Set 1 with $k_\mathrm{max}=1$ corresponds to an error of $\epsilon_\mathrm{max} = 20\%$, and when considering non-heterogeneous continuous beam systems (reflected by Set 2, 3 and 4), this error can increase up to $\epsilon_\mathrm{max} = 50\%$ and more. This is further evidence, as already pointed out in literature, that the load arrangement provisions from building codes can be non-conservative and hence lead to unsafe designs \cite{mullen_bounds_1999}.

The simplified provisions within the Eurocodes, which also exist within EN 1992-1-1 5.1.3 \cite{CEN_2015_BS_199211_Eurocode} and other codes \cite{TechnicalCommittee_2020_Code_of_Practice}, need to be understood in context of the $1.5 Q_k$ load factors and the dead load contribution $G_k$, which invariably will lessen the underestimation made by the provisions. Nonetheless, the validity of the design code recommendations for flexural load arrangements could be investigated further, especially for highly irregular beam and floor arrangements \cite{ho_pattern_2015}.

\subsection{Shear load arrangements}
Unlike flexural load arrangements, which have been identified in literature and building codes, the shear load arrangements are a novel discovery. To the authors' knowledge, this is the first time that deep beams have been identified to cause new critical load arrangements in literature. Although shear load arrangements sometimes resulted in identical utilisation ratios to that of flexural ones, initial analyses pointed to an average increase in utilisation ratio of 4-5\%, while larger deviations were occasionally observed. Figure \ref{fig:validation_algorithms} also highlights that these shear load arrangements were relatively prevalent within the design scenarios considered.

Confirmation and validation of these shear load arrangements by future research is encouraged. Of particular interest is why Equation \ref{eq:max_shear_beam_span} defines the exact point when these critical load arrangements arise. One notable difference in the mechanical assumption in this investigation of load arrangements as to that of previous studies was the use of Timoshenko-Ehrenfest rather than Euler–Bernoulli beam theory. For example, the two seminal works on establishing the bounds of critical load arrangements using fuzzy set based finite-element methods used Bernoulli-Euler beam theory \cite{koyluoglu_interval_1995, mullen_bounds_1999}. A re-investigation with deep beams as defined by Equation \ref{eq:shear_beam_span} and Timoshenko-Ehrenfest beam theory should reveal more critical bounds of load arrangements than previously identified with interval-finite-element methods. The extent to which these shear load arrangements require special provisions within building codes will require further exploration.

\subsection{Critical load arrangement algorithms}

The critical \textit{load arrangement algorithms} provided in \ref{app:algorithm_1} and \ref{app:algorithm_2}, along with a study of their computational complexity, were key for the evaluation of the influence zone. Limiting the design space to a fraction of the naive $J_{naive}$ load arrangement set without making heuristic simplifications was crucial in both the data set generation and influence zone evaluation steps.

It is likely that there is further room for improvement for Algorithm 2 for evaluating the shear load arrangements for a known list of susceptible shear beams. The current formulation, as explained in Section \ref{sec:shear_load_arrangements}, still leads to either pre-existing flexural load arrangements, or creates duplicate shear load arrangements. On average, 74.7\% of the outputs obtained from Algorithm \ref{app:algorithm_2} were unique, with a best-case efficiency of 88.8\% and a worst-case efficiency of 12.7\%. This suggests that an algorithm with a lesser computational complexity than $O(m^2\, 2^n)$ might be achievable through further investigation.

\subsection{Future investigations and application of the influence zone concept}

This investigation will hopefully serve as a starting point for future studies related to the influence zone. There were several limitations within this study, notably not accounting for serviceability checks and limiting the design space to positively loaded UDLs. Furthermore, only 11 equidistant points were sampled about each beam during design and extraction of influence zone values. A more efficient approach could sample specific points against the critical load arrangement that apply to that particular polarity zone.

Further studies could be conducted for different material and design information assumptions, while studies could also be expanded to 2D continuous frames and shells, with fixed and semi-rigid connections. As previously explained, such numerical studies could be validated with either analytical or experimental approaches, along with using local, as opposed to global, influence zone formulations as discussed in section \ref{sec:influence_zone_formulation}.

The influence zone concept and associated results could be a helpful piece of information when teaching the design of large-scale, steel-framed continuous beam systems, may have applications in other research areas such as reliability engineering, and help in the future development of generalised design models \cite{gallet_structural_2022}.

%% file: 06_Conclusion.tex
\section{Conclusions}
\label{sec:conclusion}

A novel concept termed the \textit{influence zone} was proposed in relation to continuous beam systems. The investigation developed a local and global formulation, of which the latter one was explored numerically with design constraints applicable to steel framed buildings. The key challenge was the explicit definition of critical load arrangements to allow the computational feasible generation of design data sets and evaluation of their respective influence zones. The investigation led to three important outcomes:
\begin{itemize}
    \item The development of polarity sequences and polarity zones which led to the demarcation between previously known flexural load arrangements and the newly discovered shear load arrangements, with an explicit span limit equation for when these novel load arrangements occur.
    \item Two algorithms capable of finding these two types of load arrangements, and providing evidence that they encompass all critical permutations in comparison to the naive, brute-force approach.
    \item The generation of design data sets from which the influence zone values for various degrees of design complexities and error thresholds could be rigorously studied. For error thresholds deemed acceptable in structural design, the influence zone for continuous beams within steel framed building under ultimate state considerations is on average less than 3, going to a maximum influence zone value of 5.
\end{itemize}

The influence zone is a heuristic design tool that differentiates itself from influence lines (and influence surfaces) and demonstrates the value of the inverse problem perspective through which it was evaluated by. This study opens the scope for future research, notably in the evaluation of influence zones for various materials and structural systems, validating and explicating the existence of shear load arrangements, and encouraging research on improving the existing algorithm that identifies them. 